\documentstyle[aps,graphicx,floats,twocolumn]{revtex}

\begin{document}

\draft   

\wideabs
{
\title{Strong Electron-Phonon Coupling in Superconducting MgB$_2$: A
Specific Heat Study}

\author{Ch. W\"alti$^{1}$, E. Felder$^{1}$, C. Degen$^{1}$, G. Wigger$^{1}$, 
R. Monnier$^{1}$, B. Delley$^{2}$ and H. R. Ott$^{1}$}
\address{$^{1}$Laboratorium f\"ur Festk\"orperphysik, ETH-H\"onggerberg, 
         8093 Z\"urich, Switzerland}
\address{$^{2}$PSI, 5232 Villigen, Switzerland}

\maketitle

\begin{abstract}
We report on measurements of the specific heat of the recently discovered
superconductor MgB$_2$ in the temperature range between 3 and 220~K. 
Based on a modified Debye-Einstein model, we have achieved a rather
accurate account of the lattice contribution to the specific heat, which
allows us to separate the electronic contribution from the total measured
specific heat. From our result for the electronic specific heat, we estimate
the electron-phonon coupling constant $\lambda$ to be of the order of 
2, significantly enhanced compared to common weak-coupling values 
$\leq 0.4$.
Our data also indicate that the electronic specific heat in the superconducting
state of MgB$_2$ can be accounted for by a conventional, $s$-wave type BCS-model.
\end{abstract}

}

The recent discovery of superconductivity in MgB$_2$ below $T_c \approx
39$~K~\cite{Akimitsu} has caused a remarkable excitement in the solid-state
physics community. 
Critical temperatures of this magnitude inevitably raise the 
question whether mechanisms other than the common electron-phonon 
interaction are responsible for the transition.
In their very recent work, Bud'ko {\em et al.}~\cite{2569} have
investigated the Boron isotope effect in superconducting MgB$_2$ and found that
replacing $^{11}$B by $^{10}$B increases the critical temperature by about 1~K.
This was taken as strong evidence that superconductivity in MgB$_2$ is of
conventional nature, i.e., the electron pairing interaction is
phonon-mediated. Kotagawa {\em et al.} have interpreted 
their $^{11}$B NMR 
measurements as indicating strong coupling $s$-wave 
superconductivity~\cite{2570}. Preliminary $^{11}$B NMR measurements 
at low temperatures yield spectra which are consistent with the 
expectations for a type-II superconductor in the mixed 
state~\cite{Jorge}. Evidence for sizeable electron-phonon coupling is 
also provided by recent band-structure 
calculations~\cite{2572,2584,2585}.
Various tunneling experiments have provided some evidence for 
conventional BCS-like superconductivity, but the values of the 
superconducting energy gap extracted from these measurements vary from 
2 to 7 meV~\cite{2574,2575,2576}.

In this letter, we report on measurements of the specific heat 
$C_{p}$ of MgB$_2$ and present $C_{p}(T)$ data in a wide 
temperature range.
Using a consistent model for the contribution of the lattice 
vibrations to $C_{p}(T)$ we calculate
the electronic specific heat of this material.
We show that the electronic specific heat below $T_c$ is in good agreement
with a conventional BCS-type interpretation.

The sample has been prepared from commercially available 
MgB$_{2}$-powder (Alfa Aesar) by sintering a pressed pellet at 500$^{\circ}$C for 
about 72~h. Electron microprobe investigations of the sample have shown that  
impurities of heavy elements (Cu, Ni, W) with concentrations of the order of $10^{-2}$ 
are present in the sample.

The specific heat $C_p(T)$ of the sintered MgB$_2$ sample has been measured
using two different experimental techniques in overlapping temperature ranges.
A standard relaxation technique was employed in the temperature range between 3
and 45 K. For temperatures between 20 and 220 K an adiabatic continuous heating
calorimeter was used. Special care was taken to minimize the radiation losses
at elevated temperatures~\cite{2444}. The temperatures in the range between
3 and 45~K were reached using a pumped $^4$He~cryostat, and for those between
20 and 220~K, a conventional gas-flow $^4$He~cryostat was used.

\begin{figure}[ht]
\begin{center}\leavevmode
\includegraphics[width=0.9\linewidth]{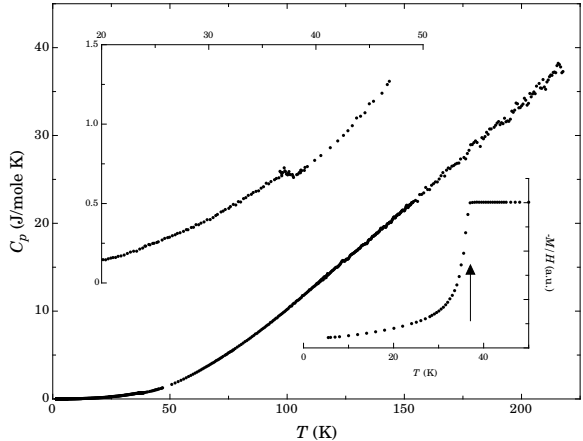}
\caption{Specific heat $C_p(T)$ of sintered MgB$_2$ as a function of temperature
between 3 and 220~K. The upper inset shows the same quantity in a 
limited
temperature range. The lower inset displays the magnetisation $M$ divided by 
the applied field $H$ as a function of temperature
for our sample. The vertical arrow marks the onset of superconductivity in
this material at $T_c \approx 37.5$~K.
}\label{Fig:Cp-tot}\end{center}\end{figure}

In the lower inset of Fig.~\ref{Fig:Cp-tot} we show the magnetisation 
of our sample divided by the constant applied field,
$M/H$, as a function of temperature for $T<50$~K. 
The superconducting transition
temperature is 37.5~K and is indicated by the vertical
arrow in the figure. The reduction of $T_c$ of our sample, compared to 
$T_c \approx
39$~K observed by Bud'ko and coworkers~\cite{2569}, is most likely due to
the fact that our sample is not as clean as theirs and the superconducting
transition temperature is slightly reduced by the impurities. We note,
however, that the anomaly in $C_{p}(T)$ at $T_{c}$ is at least as sharp as the 
one reported in Ref.~\onlinecite{2569}.

In Fig.~\ref{Fig:Cp-tot} we show the as measured specific heat $C_{p}(T)$ vs. $T$ in
the whole temperature range covered in this study. The upper inset of this figure
shows the same data in a limited temperature range around $T_{c}$. The anomaly
in the specific heat due to the superconducting transition of MgB$_2$ at about
37.5~K is clearly resolved. We note that the absolute values of the data
presented here are in good agreement with previously reported results of the specific
heat of MgB$_2$ in a narrow range of temperatures around $T_c$~\cite{2569}. 

With respect to superconductivity in MgB$_{2}$, 
the most interesting part of $C_{p}(T)$ is the electronic specific heat
$C_{el}(T)$. In order to reliably separate this
contribution from the total, measured specific heat, the contribution due to
lattice excitations, $C_{ph}(T)$, has to be known quite accurately. 
As may be seen in Fig.~\ref{Fig:Cp-tot}, the lattice provides the 
dominant
contribution to the total, measured specific heat above 20~K. 
Well below the Debye-temperature $\theta_{D}$, the specific heat in the normal state
of a common metal is usually approximated
by $C_p(T) = \gamma T + \beta T^3$, where the first
term represents the electronic and the second term the lattice specific heat.
We show below that this approximation for evaluating both, the electronic 
and the lattice specific heat, is not applicable for MgB$_2$ at $T 
\geq T_{c}$. Even at temperatures only slightly above $T_c$, the lattice specific
heat may not simply be described by assuming a linear dispersion of the acoustic modes. 

\begin{figure}[ht]
\begin{center}\leavevmode
\includegraphics[width=0.9\linewidth]{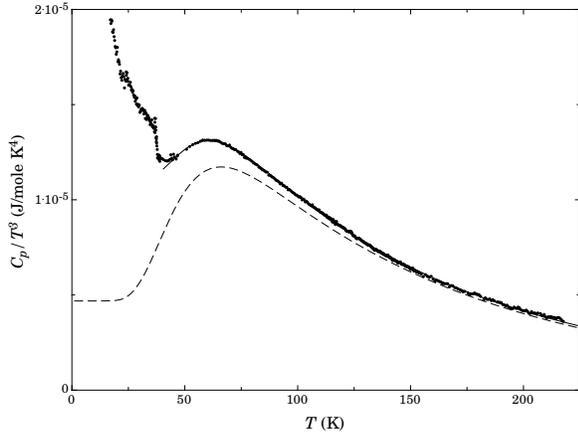}
\caption{Specific heat $C_p(T)$ of MgB$_2$ plotted as $C_p(T)/T^3$. The solid line
represents a fit of Eq.~(\protect\ref{Eq:Cp-Fit}) to the data
at $T>T_c$, and the broken line represents the resulting lattice contribution 
(see text).
}\label{Fig:Cp-ph}\end{center}\end{figure}

In Fig.~\ref{Fig:Cp-ph}, we show the total measured specific heat of MgB$_2$
divided by $T^3$ as a function of $T$. We note that, just above $T_c$, $C_p(T)/T^3$ 
increases with increasing temperature  and passes through a 
pronounced local 
maximum at about 60~K. Such a feature cannot be described by the 
approximation $C_p(T) = \gamma T + \beta T^3$, mentioned above.

Due to the present lack of thermal expansion data, we cannot calculate 
the specific heat $C_V(T)$ at
constant volume. The difference between $C_p(T)$ and $C_V(T)$ is 
expected to be small in the entire covered temperature range and at 
this point we neglect it.

We assume that the total specific heat at $T>T_c$ may be described by an
electronic contribution $C_{el}(T)$ which is given by $\gamma T$ at 
$T>T_c$ in the whole
covered temperature range, and a lattice contribution $C_{ph}(T)$. 

The lattice specific heat is generally given by
\begin{equation}
C_{ph}(T)= \int_{0}^{\infty} d\omega g(\omega) \frac{\hbar^2 \omega^2}{k_B T^2}
\frac{\exp(\frac{\hbar \omega}{k_B T})}{\left(\exp(\frac{\hbar \omega}{k_B
T})-1\right)^2} \, ,  
\end{equation}
where $g(\omega)$ denotes the phonon density of states (PDOS), $\hbar$ the
Planck and $k_B$ the Boltzmann constant, respectively. In the Debye
approximation, the lattice is treated as an isotropic continuum with a
linear dispersion, which leads to a PDOS proportional to $\omega^2$ for
$\omega < \omega_D$,  $\omega_D$ denoting the cut-off frequency above
which the PDOS is zero, and accordingly, to a low-temperature lattice specific
heat of the form $C_{ph}(T)/T^3 = $constant. As mentioned above, it is not
possible to describe the observed maximum in $C_p(T)/T^3$ by this simple
model. The Debye model may easily be extended to include deviations from the
linear dispersion. 
This leads to a PDOS of the form $g(\omega) = \mu \omega^2
+ \nu \omega^4$ and, in the low temperature limit, to $C_{ph}/T^3= \beta +
\delta T^2$. The cut-off frequency $\omega_D$ is chosen such that the total
PDOS per mole is limited to $3pN_A$, where $p$ is the number of atoms per unit
cell and $N_A$ Avogadro's constant. The lattice specific heat in this
extended Debye-scheme is then given by 
\begin{equation} \label{Eq:Cp-Debye}
C_{D}(T)= \int_{0}^{\omega_D} d\omega (\mu \omega^2 + \nu
\omega^4) \frac{\hbar^2 \omega^2}{k_B T^2} \frac{\exp(\frac{\hbar \omega}{k_B
T})}{\left(\exp(\frac{\hbar \omega}{k_B T})-1\right)^2} \, .  
\end{equation}

Kortus and coworkers~\cite{2572} have recently computed the energies 
of the zone-center optical modes in MgB$_2$ using a frozen phonon 
scheme. According to their work, the
lowest optical mode is doubly degenerate and located around an energy of
$\hbar \omega_{opt} / k_{B} \approx 460$~K. 
A calculation of the phonon spectrum yields a peak in the PDOS
at $\hbar \overline{\omega}_{opt} / k_{B} \approx 380$~K~\cite{2584}.
Modes of this energy 
contribute to the lattice specific heat already at temperatures far below $T_c$
and thus cannot be neglected in our analysis. 
In order to take them into account,
we include their contribution into our approximation of the lattice specific
heat by treating them as Einstein modes. The specific heat of one
Einstein-mode ($N_A$ states at an energy $\hbar \omega_E$) is given by
\begin{equation} \label{Eq:Cp-Einstein}
  C_{opt}(T) = N_A \frac{\hbar^2 \omega_E^2}{k_B T^2} \frac{\exp(\Theta_E / T
)}{ \left( \exp(\Theta_E / T ) -1 \right)^2 } \, , 
\end{equation}
where $\Theta_E = \frac{\hbar \omega_E}{k_B}$ denotes the
Einstein-temperature, and $\hbar \omega_E$ the energy of the optical mode.

MgB$_2$ has $p=3$ atoms per unit cell, and therefore 9 phonon-modes. We treat
the energetically lowest optical mode, which is doubly degenerate, 
according to Eq.~(\ref{Eq:Cp-Einstein}) and the remaining 7 modes by the
formula given in Eq.~(\ref{Eq:Cp-Debye}). The solid line in
Fig.~\ref{Fig:Cp-ph} is a fit based on
\begin{equation} \label{Eq:Cp-Fit} 
C_p(T)/T^3 =\gamma/T^{2} + C_D(T)/T^3 + C_{opt}(T)/T^3 
\end{equation} 
to the data at $T>T_c$. By inspecting Fig.~\ref{Fig:Cp-ph}, it may be seen that
with this model the measured total specific heat $C_p(T)$
may be well approximated at $T>T_{c}$, thus 
providing a rather accurate description of the lattice and the electronic
specific heat of MgB$_2$ in its normal state. 

The fitting procedure provides a value  of 
$5.5 \; \mathrm{mJ}/\mathrm{mole \cdot K^2}$ for 
$\gamma$. This value is almost twice as large as the
$\gamma$ value given in Ref.~\cite{2569}. However, considering the high accuracy 
of our fit we are confident that our value for $\gamma$ is quite reliable. 
If we interpret the $\gamma$ parameter as the Sommerfeld constant, a 
comparison with recently calculated
densities of states at the Fermi-energy $E_F$, $D(E_F) = 
0.72$~states/unit-cell$\cdot$eV~\cite{2572} and 0.74
states/unit-cell$\cdot$eV~\cite{Rene}, respectively, leads directly 
to a value $\frac{m^{\ast}}{m} = 3.14$ for the average mass-enhancement 
of the conduction electrons in MgB$_2$. Neglecting other many-body 
effects, the corresponding electron-phonon coupling constant $\lambda =
\frac{m^{\ast}}{m}-1 = 2.14$ is significantly enhanced above the
usual BCS-weak-coupling values of $\lambda_{wc} < 0.4$, thus providing evidence for
a considerable electron-phonon coupling in MgB$_2$. 

The lattice contribution to the specific heat extracted from the fit is shown
in Fig.~\ref{Fig:Cp-ph} as a broken line. Since our model reproduces 
the total specific heat at elevated temperatures 
rather well, we may safely assume that our calculation provides the lattice
contribution not only for $T>T_c$, but across the whole covered temperature
range. The other two main parameters which emerge from the fit are the Debye
temperature $\theta_D = 746$~K~\cite{ThetaD}, in good agreement with 
the value given in Ref.~\cite{2569}, and the Einstein temperature 
$\theta_E=325$~K.
The corresponding energy for this dispersionless mode is in  
reasonable agreement 
with the  position of the peak in the PDOS calculated in 
Ref.~\cite{2584} and 
observed at $\hbar \omega / k_{B} = 365$~K in the inelastic neutron scattering data of 
Ref.~\cite{2580}.

\begin{figure}[ht]
\begin{center}\leavevmode
\includegraphics[width=0.9\linewidth]{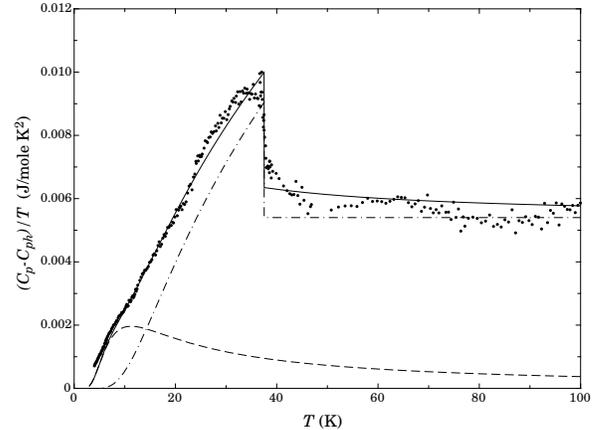}
\caption{Total specific heat $C_p(T)$ minus the lattice contibution $C_{ph}(T)$ divided
by $T$ vs. temperature. The solid line represents a fit to our data 
using Eq.~(\protect\ref{Eq:Cp-Fit-Res}). The broken line represents the
resonance contribution due to the impurities and the dash-dotted line the
calculated electronic specific heat (see text). 
}\label{Fig:Cp-el-res}\end{center}\end{figure}
With the proviso that the lattice specific heat $C_{ph}(T)$ is now 
established at all covered temperatures, we calculated the
electronic specific heat of MgB$_2$ by subtracting $C_{ph}(T)$ from 
the total measured specific heat. The result of this
calculation is shown in Fig.~\ref{Fig:Cp-el-res} by the closed circles.

As mentioned in the introduction it seems quite likely that superconductivity in MgB$_2$ is
driven by conventional electron-phonon coupling. 
It has been shown that the specific heat of conventional, 
electron-phonon coupling driven superconductors is, regardless of the 
coupling-strength, well described by the usual BCS-expression, but 
scaled by the factor $\alpha_{s}=\frac{\Delta(0)}{1.76 k_{B} 
T_{c}}$~\cite{2579}.
Therefore, the
electronic specific heat at $T<T_c$ may be written as~\cite{2579,Parks}
\begin{eqnarray} \label{Eq:Cp-BCS}
C_s(T) = \left( \frac{\Delta(0)}{1.76 k_{B} T_{c}}\right) 
              \left(- \frac{D(E_F) (1+\lambda) }{T} \right) \times \nonumber \\
	     \times \int_{-\infty}^{\infty} d\epsilon 
		 \left(\epsilon^2 +\tilde{\Delta}^2 + \frac{1}{2 k_B T} 
		 \frac{\partial \tilde{\Delta}^2}{\partial
                   (k_B T)^{-1}} \right) \frac{\partial f}{\partial E}, 
\end{eqnarray}
where $\Delta(0)$ denotes the superconducting energy-gap at $T=0$~K
and $\tilde{\Delta}$ is the temperature dependent BCS gap-function.
This function has been tabulated by M\"uhlschlegel~\cite{2573}.
The numerical evaluation of Eq.~(\ref{Eq:Cp-BCS}) is presented as 
the dash-dotted line in Fig.~\ref{Fig:Cp-el-res}.
We note that our data are not well reproduced in this way, 
but, as we show below, the apparent extra contribution is due to lattice
excitations of the impurities.

As already mentioned above, our sample contains impurities at concentrations of
the order of 1\%. Since Boron and also Magnesium are rather light atoms, 
the identified
impurity atoms are certainly much heavier. Heavy impurity atoms may cause
resonances in the continuum of the phonon excitation spectrum at energies
well below the energy of the lowest optical mode~\cite{2577}. In order
to separate such contributions to the specific heat from the electronic
specific heat, we fitted our data with an expression of the form
\begin{equation} \label{Eq:Cp-Fit-Res}
C_p(T)-C_{ph}(T) = C_{el}(T) + C_{res}(T) \, , 
\end{equation}
where $C_{el}(T)$ for
$T<T_c$ is given by Eq.~(\ref{Eq:Cp-BCS}) and for $T>T_c$ by $\gamma T$. 
To keep the model as simple as possible, the
contribution to the specific heat due to the resonant modes, $C_{res}(T)$, is
taken into account by an Einstein term similar to Eq.~(\ref{Eq:Cp-Einstein}).
The impurity concentration $n_{imp}$ enters
as an additional free parameter. The fit gives 
$n_{imp}=0.5$\%, in line with our expectations, a resonance
energy of 30~K, and $\Delta(0) = 1.2 k_{B} T_{c}$. The ratio 
$\Delta(0) / k_{B} T_{c}$ is surprisingly small but, as has been shown by Swihart~\cite{2578}, 
even for substantially enhanced electron-phonon couplings, $\frac{\Delta(0)}{k_{B} 
T_{c}}$ may be reduced to below the original weak-coupling value of 
1.76. This is particularly the case if low-energy dispersionless 
phonon modes are present. 

The presence of impurities does not significantly alter the
total specific heat at elevated temperatures. Their contribution to the lattice 
specific heat above $T_{c}$ is
less than $10^{-3}$ and can safely be
neglected in the calculation described above in fitting the specific heat at $T>T_c$. 

\begin{figure}[ht]
\begin{center}\leavevmode
\includegraphics[width=0.9\linewidth]{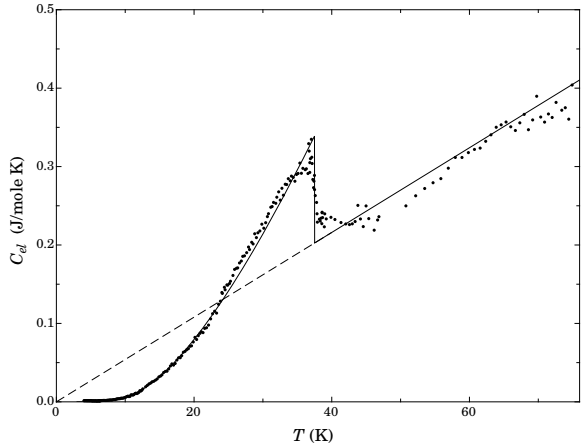}
\caption{Electronic specific heat $C_{el}(T)$ of MgB$_2$ as a function of temperature.
The solid line represents the rescaled BCS-expectation of the electronic specific
heat (Eq.~(\ref{Eq:Cp-BCS})), and the broken line is the 
electronic specific heat in the hypothetical normal-state.
}\label{Fig:Cp-el}\end{center}\end{figure}
In Fig.~\ref{Fig:Cp-el} we show the electronic contribution to the specific heat
in the temperature range between 3 and 75~K, which is extracted from the
total specific heat by subtracting the lattice contribution, including the
small resonant term discussed above. The solid line in this figure is 
calculated using Eq.~(\ref{Eq:Cp-BCS}). We note a rather good
agreement between the calculations and our data, which
provides further evidence that superconductivity in MgB$_2$ is well
described by the BCS approximation.

In conclusion, we have presented a detailed analysis of experimental 
specific heat data for MgB$_{2}$, covering an extended range of 
temperatures. The lattice specific heat is compatible with a  
Debye-temperature $\theta_D= 746$~K. 
The electronic specific heat is rather well described by the BCS approximation, 
assuming a zero-temperature energy-gap of $\Delta(0) = 1.2 k_{B} T_{c}$.
Finally, we have obtained an electron-phonon
coupling constant $\lambda \approx 2$, which is distinctly larger than
the usual weak-coupling values.
This value of $\lambda$ should be considered as an upper limit. If 
the difference $C_{p}-C_{V}$, approximately linear in 
$T$, cannot be neglected, this would automatically lead to a 
reduction of $\lambda$. Indeed, with the published value for the bulk 
modulus $B=151$~GPa~\cite{2582}, and a reasonable value for the volume
thermal expansion coefficient $b=2.5 \times 10^{-5}$~K$^{-1}$ at 
$T>T_{c}$, we obtain $(C_{p}-C_{V})/T = 1.67 \; \mathrm{mJ}/\mathrm{mole \cdot K^2}$.
This would lead to a reduction of the electronic specific heat 
coefficient to $\gamma = 3.83 \; \mathrm{mJ}/\mathrm{mole \cdot K^2}$ and 
a corresponding reduction of $\lambda$ to a value of 1.18. According 
to Eq.~(\ref{Eq:Cp-BCS}), the ratio $\Delta(0)/k_{B} T_{c}$ would 
thus be 
enhanced to a value of 1.73.

The samples used in this work were prepared by S. Sigrist. We acknowledge 
the microprobe analysis investigations by P. W\"agli. This work was in part
financially supported by the Schweizerische Nationalfonds zur F\"orderung der
wissenschaftlichen Forschung.


\begin{thebibliography}{10}

\bibitem{Akimitsu}
J. Akimitsu, Symposium on Transition Metal Oxides, Sendai  (2001).

\bibitem{2569}
S.~L. Bud'ko, G. Lapertot, C. Petrovic, C.~E. Cunningham, N. Anderson, and
  P.~C. Canfield, cond-mat/0101463  (2001).

\bibitem{2570}
H. Kotegawa, K. Ishida, Y. Kitaoka, T. Muranaka, and J. Akimitsu,
  cond-mat/0102334  (2001).

\bibitem{Jorge}
J.~L. Gavilano, D. Rau, Sh. Mushkolaj, and H.~R. Ott, private communication
  (2001).

\bibitem{2572}
J. Kortus, I.~I. Mazin, K.~D. Belashchenko, V.~P. Antropov, and L.~L. Boyer,
  cond-mat/0101446  (2001).

\bibitem{2584}
Y. Kong, O.~V. Dolgov, O. Jepsen, and O.~K. Andersen, cond-mat/0102499  (2001).

\bibitem{2585}
J.~M. An and W.~E. Pickett, cond-mat/0102391  (2001).

\bibitem{2574}
G. Rubio-Bollinger, H. Suderow, and S. Vieira, cond-mat/0102242  (2001).

\bibitem{2575}
H. Schmidt, J.~F. Zasadzinski, K.~E. Gray, and D.~G. Hinks, cond-mat/0102389
  (2001).

\bibitem{2576}
A. Sharoni, I. Felner, and O. Millo, cond-mat/0102325  (2001).

\bibitem{2444}
O. Jeandupeux, Ph.D. thesis, ETH Z\"urich, 1996.

\bibitem{Rene}
B. Delley and R. Monnier (unpublished).

\bibitem{ThetaD}
In our approach, $\Theta_D$ is a free fit-parameter entering the parameter
  $\mu$ in Eq.~(\protect\ref{Eq:Cp-Debye}).

\bibitem{2580}
T.~J. Sato, K. Shibata, and Y. Takano, cond-mat/0102468  (2001).

\bibitem{2579}
H. Padamsee, J.~E. Neighbor, and C.~A. Shiffman, J. Low Temp. Phys. {\bf 12},
  387  (1973).

\bibitem{Parks}
G. Rickayzen,  in {\em Superconductivity}, edited by R.~D. Parks (Marcel
  Dekker, inc., New York, 1969), p.\ 79.

\bibitem{2573}
B. M\"uhlschlegel, Z. Phys. {\bf 155},  313  (1959).

\bibitem{2577}
C.~T. Walker and R.~O. Pohl, Phys. Rev. {\bf 131},  1433  (1963).

\bibitem{2578}
J.~C. Swihart, Phys. Rev. {\bf 131},  73  (1963).

\bibitem{2582}
T. Vogt, G. Schneider, J.~A. Hriljac, G. Yang, and J.~S. Abell,
  cond-mat/0102480  (2001).

\end{thebibliography}


\end{document}